\documentclass[reprint,amsmath,amssymb,aps,prb,showpacs]{revtex4-2}
\usepackage{lineno}
\usepackage[english]{babel}
\usepackage{multirow}
\usepackage{bm}
\usepackage{xspace}
\usepackage{ulem}
\usepackage{soul}
\usepackage{siunitx}
\usepackage{graphicx}% Include figure files
\graphicspath{{figures//}} %allows to omit the path in the \incudegraphics command
\usepackage{dcolumn}% Align table columns on decimal point
\usepackage{bm}% bold math
\usepackage{color}
\usepackage{wrapfig}
\usepackage{hyperref}
\usepackage{array}
\usepackage{booktabs}
\usepackage{multirow}
\usepackage{amssymb}

\usepackage{marvosym}
\hypersetup{
    unicode=false,          % non-Latin characters in Acrobat’s bookmarks
    pdftoolbar=true,        % show Acrobat’s toolbar?
    pdfmenubar=true,        % show Acrobat’s menu?
    pdffitwindow=false,     % window fit to page when opened
    pdfstartview={FitH},    % fits the width of the page to the window
    pdftitle={My title},    % title
    pdfauthor={Author},     % author
    pdfsubject={Subject},   % subject of the document
    pdfcreator={Creator},   % creator of the document
    pdfproducer={Producer}, % producer of the document
    pdfkeywords={keyword1} {key2} {key3}, % list of keywords
    pdfnewwindow=true,      % links in new window
    colorlinks=true,       % false: boxed links; true: colored links
    linkcolor=blue,          % color of internal links (change box color with linkbordercolor)
    citecolor=blue,        % color of links to bibliography
    filecolor=blue,      % color of file links
    urlcolor=blue,       % color of external links
    plainpages=false        % influences the page numbering
}

\newcommand{\rev}[1]{\textcolor{red}{#1}}
\let\rev\relax

\begin{document}

\title{Fractional phase slips across the charge-density-wave domain walls in 1-$T$ TiSe$_2$}
\author{Haotian Zhang$^{1,*}$}
\author{Zihao Song$^{1,2,*}$}
\author{Zhongchen Xu$^{3,*}$}
\author{Jun Shu$^{1}$}
\author{Zhongxu Wei$^{3}$}
\author{Zunming Lu$^{2}$}
\author{Jun Liu$^{1}$}
\author{Zengyi Du$^{4}$}
\author{Jinxing Zhang$^{5,8}$\textrm{\Letter}}
\author{Youguo Shi$^{3,6}$\textrm{\Letter}}
\author{Ge He$^{1,7}$\textrm{\Letter}}
\author{Jun Shen$^{1}$\textrm{\Letter}}

\affiliation{
$^1$ School of Mechanical Engineering\mbox{,} Beijing Institute of Technology\mbox{,} Beijing 100081\mbox{,} China \\
$^2$ School of Materials Science and Engineering, Hebei University of Technology, Tianjin 300130, China\\
$^3$ Beijing National Laboratory for Condensed Matter Physics\mbox{,} Institute of Physics\mbox{,} Chinese Academy of Sciences\mbox{,} Beijing 100190\mbox{,} China \\
$^4$ Hefei National Laboratory\mbox{,} and New Cornerstone Science Laboratory\mbox{,} Hefei, Anhui 230088\mbox{,} China\\ 
$^5$ School of Physics and Astronomy\mbox{,} Beijing Normal University\mbox{,} Beijing 100875\mbox{,} China \\
$^6$ Songshan Lake Materials Laboratory\mbox{,} Dongguan\mbox{,}Guangdong 523808\mbox{,} China\\
$^7$ Beijing Key Laboratory of Quantum Matter State Control and Ultra-Precision Measurement Technology\mbox{,} Beijing Institute of Technology\mbox{,} Beijing 100081\mbox{,} China\\
$^8$ Key Laboratory of Multiscale Spin Physics, Ministry of Education\mbox{,} Beijing 100875\mbox{,} China\\
$^{*}$ These authors contributed equally: Haotian Zhang, Zihao Song and Zhongchen Xu\\
\textrm{\Letter} e-mail: jxzhang@bnu.edu.cn; ygshi@iphy.ac.cn; ge.he@bit.edu.cn; jshen@bit.edu.cn
}

\begin{abstract}
The microscopic origin of the charge density wave (CDW) in 1\textit{T}-TiSe$_2$ remains controversial, with competing scenarios based on phonon-driven lattice instability and electronically driven excitonic correlations. Here, we combine low-temperature scanning tunneling microscopy with two-dimensional lock-in phase analysis to directly resolve the local CDW phase in real space and track its evolution across individual domain walls. In homogeneous regions, the CDW phase remains uniform; by contrast, across domain walls we uncover a robust and reproducible $2\pi/3$ phase shift that occurs collectively in all three symmetry-related CDW components. This nontrivial and correlated phase-slip configuration places stringent constraints on the order-parameter manifold and challenges the simplest purely phonon-driven commensurate lock-in picture, which would instead predict a $\pi$ phase shift. A minimal free-energy model incorporating both electron-phonon and electron-hole interactions reproduces the observed phase behavior and indicates that electronic interactions play an important role in shaping the local phase structure of the CDW order. These results establish domain walls as direct real-space probes of the microscopic interactions underlying multicomponent order and provide a general phase-resolved framework for constraining competing ordering mechanisms in correlated materials.
\end{abstract}

\maketitle

\section{Introduction}

Charge density waves (CDWs) are collective electronic ordered states characterized by a periodic modulation of the charge density accompanied by a periodic distortion of the crystal lattice \cite{PougetJeanPaul:2024,ChenChihWei:2016,GrunerGeorge:1988,MonceauPierre:2012,SoumyanarayananAnjan:2013}. As prototypical symmetry-breaking instabilities in correlated quantum materials, CDWs provide a unique platform for investigating how electronic and lattice degrees of freedom cooperate to establish long-range order. A central challenge in this context is to determine whether the primary instability originates in the lattice sector or in the electronic sector, and how this hierarchy is encoded in the free-energy landscape.

Among CDW materials, the layered transition metal dichalcogenide 1\textit{T}-TiSe$_2$ has long served as a prototypical system due to its simple crystal structure \cite{ZungerAlex:1978,DiSalvoFrankJ:1976,WilsonJohnA:1969}, well-defined transition temperature ($T_{\mathrm{CDW}} \approx 200~\mathrm{K}$) \cite{CravenRichardA:1978}, and commensurate $2a \times 2b \times 2c$ triple-$\mathbf{Q}$ order formed by three symmetry-equivalent wave vectors $\mathbf{Q}_i$ \cite{DiSalvoFrankJ:1976,TyulnevIgor:2025,GaoShang:2018}. It has been intensively studied for decades, however, the microscopic driving mechanism of the CDW transition remains under active debate, with competing scenarios invoking excitonic condensation, electron-phonon coupling, or an interplay between the two \cite{PorerMichael:2014,MonneyClaude:2016,vanWezelJasper:2010a,KanekoTatsuya:2018,KiddThomasE:2002,LianChao:2020,vanWezelJasper:2010b,vanWezelJasper:2011,HedayatHamoon:2019,KurtzFelix:2024,ZhuXuetao:2015}.

\begin{figure*}[tbp] 
  \centering 
  \includegraphics[width=0.9\textwidth]{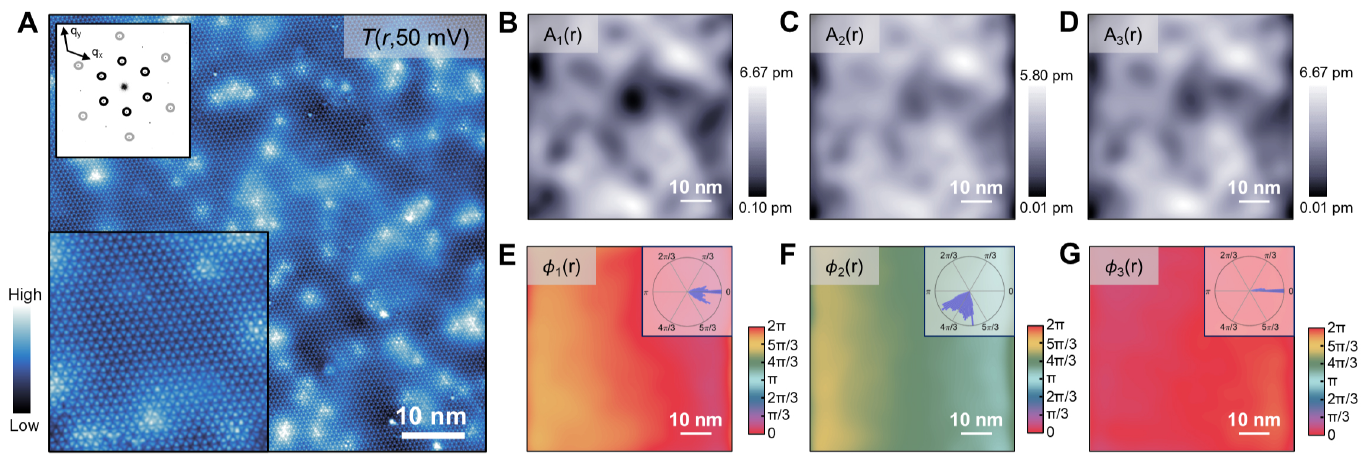} 
  \caption{ \textbf{2D lock-in analysis of the uniform triple-$\mathbf{Q}$ CDW state.} (A) STM topographic image acquired ($V = 50$~mV, $I = 1$~nA) over a $70\,\mathrm{nm} \times 70\,\mathrm{nm}$ field of view, showing both the atomic lattice corrugation and the long-wavelength CDW modulation. Upper-left inset: Two-dimensional FFT of (A), showing the Bragg peaks of the hexagonal lattice and satellite peaks located at half the reciprocal lattice vectors, confirming the commensurate $2 \times 2$ CDW order. Lower-left inset: Zoomed-in topographic image of a defect-free region. (B–D) Amplitude maps $A_i(\mathbf{r})$ obtained from 2D lock-in analysis, showing spatially homogeneous CDW amplitudes for all three symmetry-equivalent wave vectors. (E–G) Corresponding phase maps $\phi_i(\mathbf{r})$, revealing uniform phase distributions within a single CDW domain. Insets: Histograms of the phase values extracted from (E–G). $\phi_1$ and $\phi_3$ are sharply peaked near zero, whereas $\phi_2$ is centered near $4\pi/3$, demonstrating a well-defined relative phase configuration. } 
  \label{fig:topo} 
\end{figure*}

On one hand, inelastic x-ray scattering \cite{WeberFrank:2011,HoltMartin:2001},  ultrafast measurements \cite{OttoMartinR:2021,KaramTonyE:2018}, and first-principles calculations \cite{CalandraMatteo:2011,HellgrenMaria:2017,PashovDimitar:2025} have revealed phonon softening and lattice instabilities at the CDW wave vectors, supporting a scenario in which electron--phonon coupling plays a central role. However, the absence of well-defined Fermi-surface nesting and the unusually large energy scale associated with the CDW gap remain difficult to reconcile with a purely lattice-driven instability \cite{LucovskyGeorge:1977,MonneyClaude:2011}. On the other hand, angle-resolved photoemission spectroscopy (ARPES) and momentum-dependent electron energy-loss spectroscopy (M-EELS) have revealed pronounced electronic reconstruction across the CDW transition, including giant spectral-weight redistribution \cite{CercellierHerve:2007,PilloThomas:2000,SugawaraKatsuaki:2016,LinZijian:2022}, plasmon softening \cite{KogarAnshul:2017}, and ultrafast electronic response \cite{RohwerTimm:2011}, consistent with an excitonic instability between the Se $4p$ valence band and the Ti $3d$ conduction band. Rather than an either-or scenario, a cooperative mechanism involving both electron--phonon and electron-hole interactions was proposed early on by Jasper \textit{et al.} \cite{vanWezelJasper:2011}. More recently, significant theoretical and experimental advances, particularly from ultrafast spectroscopic studies \cite{PorerMichael:2014,LianChao:2020,KurtzFelix:2024}, have provided growing evidence for a cooperative interplay between lattice and electronic instabilities, leading to a self-amplified exciton--phonon mechanism \cite{LianChao:2020}. It is therefore now widely accepted that the electronic and lattice degrees of freedom in 1\textit{T}-TiSe$_2$ are strongly intertwined. The key unresolved issue is no longer whether these two sectors coexist, but which instability acts as the primary driver that reshapes the free-energy landscape at the transition.

So far, most previous experimental studies have focused on amplitude-related observables—such as gap magnitude \cite{CercellierHerve:2007,PilloThomas:2000,RossnagelKai:2002,LinZijian:2022}, phonon softening \cite{WeberFrank:2011}, incoherence effects \cite{OuYi:2024}, or lattice displacement strength \cite{DiSalvoFrankJ:1976,LianChao:2020}—such quantities predominantly probe the quadratic sector of the free-energy expansion and often remain compatible with multiple microscopic scenarios. By contrast, the internal phase configuration of the multi-$\mathbf{Q}$ order parameter directly reflects the structure and hierarchy of higher-order coupling coefficients, providing a more discriminating probe of the underlying driving mechanism \cite{LiuXiaolong:2021}. Experimental access to this phase degree of freedom in 1\textit{T}-TiSe$_2$, however, has remained comparatively limited . 

In this work, we employ state-of-the-art scanning tunneling microscopy (STM) combined with two-dimensional (2D) lock-in analysis to directly resolve the real-space phase structure of the CDW order, both in homogeneous regions and across domain walls. In homogeneous regions, the phase of each CDW component remains uniform. In contrast, across CDW domain walls, we resolve a robust $2\pi/3$ phase shift for each component. This behavior poses a challenge to a purely lattice-driven Macmillan lock-in mechanism, which would instead predict a $\pi$ phase shift across the domain wall. By incorporating both electron-phonon ($V_{e\text{-}ph}$) and electron-hole interactions ($V_{e\text{-}h}$) into a Ginzburg-Landau free-energy framework, we show that the observed phase behavior can be captured when both contributions are taken into account, with a crossover from $\pi$ to $2\pi/3$ phase shifts emerging as the relative strength of electronic interactions increases. These results indicate that purely phonon-driven commensurate lock-in descriptions are incomplete and point to a substantial electronic contribution to the CDW order in 1\textit{T}-TiSe$_2$.

Figure~1 establishes the presence of a well-developed commensurate CDW state in 1\textit{T}-TiSe$_2$ at low temperature. 
A representative STM topographic image acquired over a $70\,\mathrm{nm} \times 70\,\mathrm{nm}$ field of view (FOV) is shown in Fig.~\ref{fig:topo}A. 
As clearly seen in the lower-left inset of Fig.~\ref{fig:topo}A, in addition to the atomic-scale lattice corrugation, a pronounced long-wavelength modulation is observed uniformly across the entire area, indicative of robust CDW order. The two-dimensional fast Fourier transform (FFT) of Fig.~\ref{fig:topo}A, displayed in the upper-left inset of Fig.~\ref{fig:topo}A, exhibits sharp Bragg peaks associated with the hexagonal lattice together with distinct satellite peaks located at half the reciprocal lattice vectors. The symmetry and commensurate positions of these satellite peaks confirm the formation of a $2 \times 2$ CDW state with three symmetry-equivalent wave vectors $\mathbf{Q}_i$ in the $ab$-plane, establishing the triple-$\mathbf{Q}$ character of the ordered phase. These results are consistent with majority of previous STM work \cite{LiuChengYen:2024,NomuraAtsushi:2025,HuQiang:2024,IshiokaJunya:2010,SugawaraKatsuaki:2016,WangJingyi:2016,MulaniImrankhan:2021}.

To resolve the internal structure of the triple-$\mathbf{Q}$ order parameter, we performed a two-dimensional lock-in analysis on the topographic data shown in Fig.~\ref{fig:topo}A (see Supplementary Materials A for details). This method decomposes the total modulation into three complex components associated with the symmetry-equivalent CDW wave vectors, yielding spatially resolved amplitude $A_i(\mathbf{r})$ and phase $\phi_i(\mathbf{r})$ for each component. The extracted amplitude maps $A_i(\mathbf{r})$, shown in Figs.~\ref{fig:topo}(B-D), are spatially smooth and homogeneous across the entire FOV. No abrupt suppression or discontinuity is observed, confirming that the scanned region lies within a single CDW domain and is free of domain walls. 

The corresponding phase maps $\phi_i(\mathbf{r})$, displayed in Figs.~\ref{fig:topo}(E-G), reveal highly uniform phase distributions. To quantify the phase configuration, we compiled histograms of the phase values for each component [insets of Figs.~\ref{fig:topo}(E-G)]. $\phi_1(\textbf{r})$ and $\phi_3(\textbf{r})$ exhibit sharp distributions centered near zero, whereas $\phi_2(\textbf{r})$ shows a well-defined peak near $4\pi/3$. The narrow widths of these distributions indicate a rigid phase-locked configuration throughout the domain. This behavior is consistently reproduced in multiple domain-wall-free regions (see Supplementary Materials B). The observed phase configuration can therefore be summarized as $(\phi_1,\phi_2,\phi_3)\approx(0,4\pi/3,0)$ modulo $2\pi$. This well-defined relative phase structure establishes the intrinsic phase-locked ground state of the commensurate triple-$\mathbf{Q}$ CDW in the absence of domain boundaries and provides a reference configuration for analyzing phase evolution in spatially inhomogeneous regions.

\begin{figure*}[tbp]
  \centering
  \includegraphics[width=0.9\textwidth]{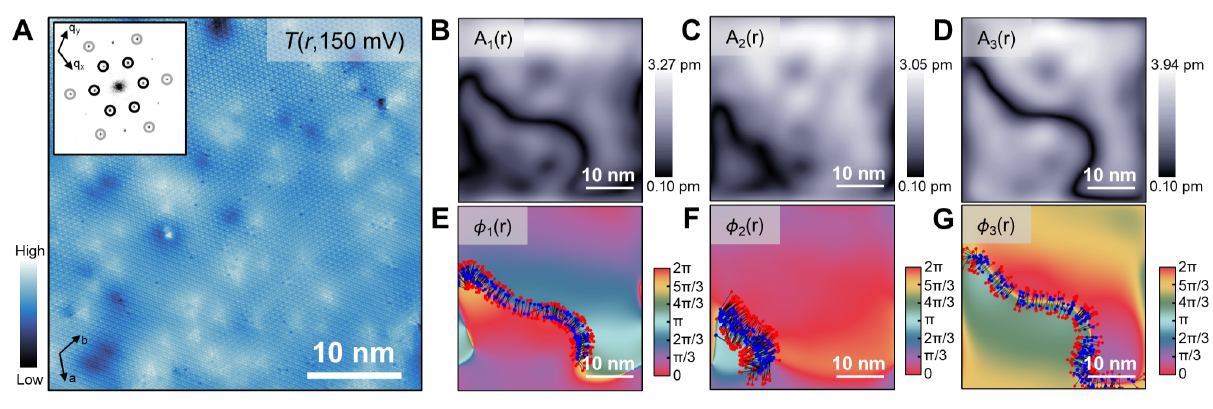}
  \caption{
  \textbf{2D lock-in analysis of CDW at the domain wall regions.}
  (A) STM topographic image acquired ($V = 150$~mV, $I = 300$~pA) over a $40\,\mathrm{nm} \times 40\,\mathrm{nm}$ field of view. Inset: FFT showing the Bragg peaks and CDW satellite peaks, confirming the persistence of commensurate order.
  (B-D) Amplitude maps $A_i(\mathbf{r})$ revealing extended domain-wall regions characterized by suppressed CDW amplitude.
  (E-G) Corresponding phase maps $\phi_i(\mathbf{r})$, showing continuous phase variation across the domain walls. Red and blue markers indicate representative sampling points located on opposite sides of the domain wall.
  }
  \label{fig:phase_cdw}
\end{figure*}

To investigate the spatial evolution of the CDW order beyond the uniform ground state, we examined regions containing domain walls. Figure~\ref{fig:phase_cdw}A shows a representative STM topographic image acquired over a $40\,\mathrm{nm} \times 40\,\mathrm{nm}$ FOV. In contrast to the uniform region, the atomic-scale contrast is clearly modified and separates into distinct regions with abrupt phase changes. The inset FFT confirms the persistence of the commensurate $2 \times 2$ CDW order, as evidenced by well-defined satellite peaks at the CDW wave vectors. Applying the same 2D lock-in procedure, we extract the amplitude maps $A_1(\mathbf{r})$, $A_2(\mathbf{r})$, and $A_3(\mathbf{r})$, shown in Figs.~\ref{fig:phase_cdw}(B-D). Unlike in the uniform region, dark boundaries with locally suppressed amplitude are clearly resolved. These features identify CDW domain walls directly in real space. The corresponding phase maps $\phi_1(\mathbf{r})$, $\phi_2(\mathbf{r})$, and $\phi_3(\mathbf{r})$ are shown in Figs.~\ref{fig:phase_cdw}(E-G). 
Across each domain wall, the CDW phase varies between adjacent domains, while remaining approximately uniform within each domain.

\begin{figure}[tbp]
  \centering
  \includegraphics[width=0.8\columnwidth]{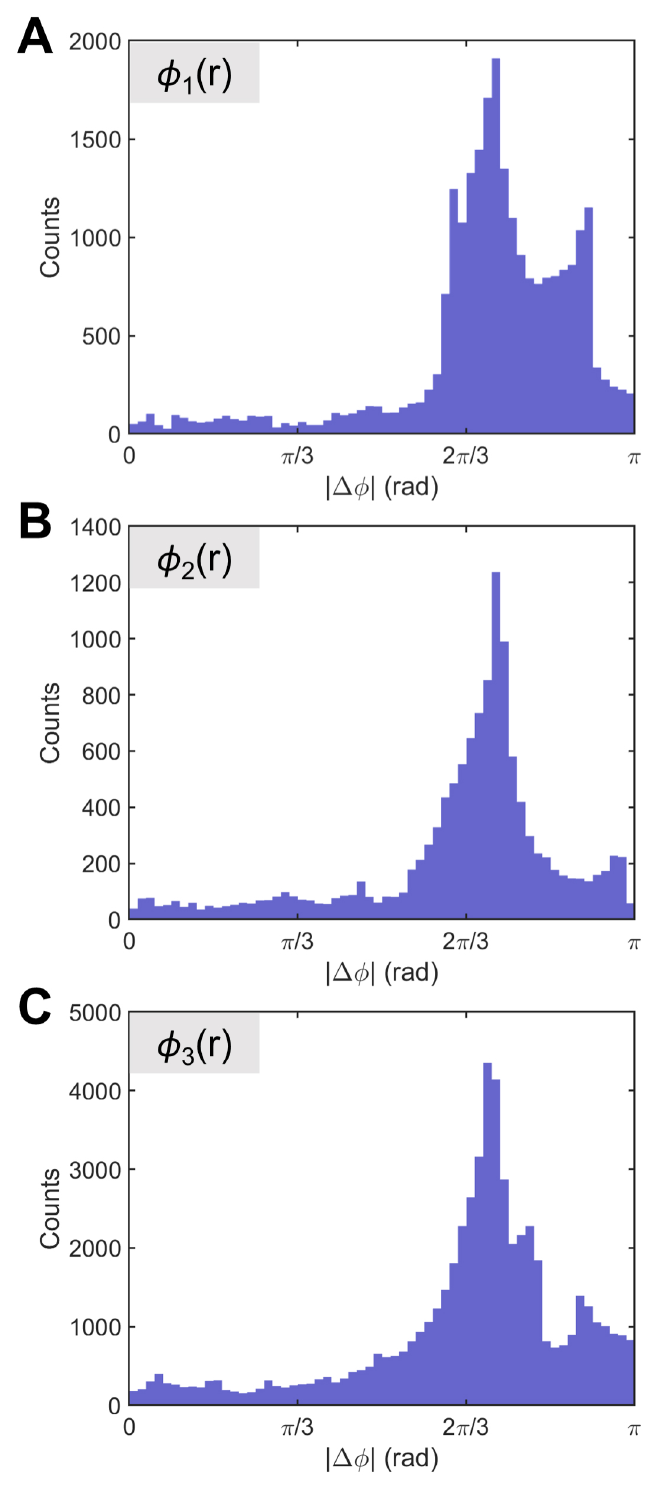}
  \caption{
  \textbf{Statistical analysis of phase differences across CDW domain walls.}
  (A-C) Histograms of the phase difference $|\Delta\phi|$ for the three CDW components, extracted across the domain walls shown in Fig.~\ref{fig:phase_cdw}(E-G). All three components exhibit well-defined peaks, indicating a robust and reproducible phase shift across the domain walls.
  }
  \label{fig:cdw_analysis}
\end{figure}

To quantify the phase difference across the domain walls, we evaluate $\Delta\phi$ by selecting pairs of sampling points located on opposite sides of the domain wall and computing the phase difference between them(see Supplementary Materials G for details of the extraction procedure). This procedure provides a local measure of the phase discontinuity across the wall. The resulting distributions of $|\Delta\phi|$ for the three CDW components are summarized in Fig.~\ref{fig:cdw_analysis}. All three components exhibit well-defined peaks, indicating a robust and reproducible phase shift across the domain walls. 

Across representative domain walls, the phase configuration evolves between distinct phase-locked states, for example from $(0,\,2\pi/3,\,4\pi/3)$ on one side to $(2\pi/3,\,0,\,0)$ on the other. Correspondingly, each CDW component undergoes a phase shift of approximately $2\pi/3$. This behavior is consistently reproduced in the other FOV (see Supplementary Materials C). The relatively narrow distributions of $|\Delta\phi|$ further indicate that the phase shift is sharply defined, rather than continuously varying, establishing a well-defined fractional phase slip across the domain walls. Such a collective yet fractional phase shift imposes a stringent experimental constraint on the microscopic origin of the CDW order.

\section{Discussion}

The phase-resolved STM measurements directly constrain the internal
phase manifold of the triple-$Q$ order parameter in
1\textit{T}-TiSe$_2$. Unlike amplitude-sensitive probes, the present
data access the relative phases of the three symmetry-related CDW
components and therefore impose quantitative restrictions on the
structure of the Ginzburg--Landau (GL) free-energy functional. To interpret the experimentally resolved phase configuration of the triple-$Q$
CDW, we first introduce a minimal Ginzburg-Landau (GL) free-energy that captures both commensurability effects and inter-component phase coupling
\cite{McMillanWilliamL:1976,vanWezelJasper:2010a}.

The GL free-energy  consistent with symmetry and commensurability
takes the form \cite{McMillanWilliamL:1976}
\begin{equation}
\label{eq:GL}
\begin{aligned}
\mathcal{F} = \int d^2\mathbf{r}\, 
& \sum_{j=1}^3 \Big[
\alpha |\Psi_j|^2
+ \frac{\beta}{2} |\Psi_j|^4
+ K |\nabla \Psi_j|^2 \Big]
\\
& + u \sum_{i<j} \left(\Psi_i \Psi_j^* + \mathrm{c.c.}\right)
+ \lambda \left(\Psi_1 \Psi_2 \Psi_3 + \mathrm{c.c.}\right),
\end{aligned}
\end{equation}

\noindent where $\Psi_j(\mathbf{r}) = A_j(\mathbf{r}) \,e^{i\phi_j(\mathbf{r})}$ is the CDW order parameter with amplitude $A_j$ and phase $\phi_j$ for each CDW component with wave vector $\mathbf{Q}_j$. The gradient term $K|\nabla \Psi_j|^2$ describes the energetic cost associated with spatial variations of the order parameter and is essential for capturing non-uniform configurations such as domain walls. The bilinear coupling $u$ correlates the relative phases between different CDW components, favoring phase alignment for $u<0$ and finite phase offsets for $u>0$. The cubic term $\lambda$ is symmetry-allowed when $\mathbf{Q}_1 + \mathbf{Q}_2 + \mathbf{Q}_3 = 0$ and couples all three CDW components simultaneously \cite{vanWezelJasper:2010a}. We emphasize that this form represents a minimal phenomenological description consistent with symmetry, and does not by itself uniquely determine the microscopic origin of the coupling terms.

The phase configuration identified in Fig.~\ref{fig:topo} corresponds to a spatially uniform CDW state, where both amplitudes and phases are
approximately constant on the scale of the STM FOV. In this limit, the amplitudes may be taken as
equal, $A_j = A$. The phase-dependent part of the GL free energy then reduces to
\begin{equation}
\mathcal{F}_{\mathrm{phase}} =
2\lambda A^3 \cos(\phi_1 + \phi_2 + \phi_3)
+ u A^2 \sum_{i<j} \cos(\phi_i - \phi_j),
\label{eq:F_phase}
\end{equation}
which provides a minimal phenomenological framework for interpreting the uniform phase
relations extracted from the STM measurements.
\\    

The phase-resolved STM analysis in Fig.~\ref{fig:topo} reveals a uniform phase configuration in which one CDW component is shifted by approximately $2\pi/3$ relative to the other two, up to symmetry-related permutations and the phase convention of the lock-in analysis. We note that while the absolute values of the extracted phases depend on the choice of reference and analysis procedure, the relative phase differences are physically meaningful and remain robust.

Such a fractional phase relation is difficult to reconcile with independent commensurate pinning of the three components, which would constrain each $\phi_j$ to either $0$ or $\pi$ \cite{McMillanWilliamL:1976,YanShichao:2017,SperaMarcello:2019}. Instead, it points to the importance of coupling between different CDW components, as captured by Eq.~(\ref{eq:F_phase}) \cite{vanWezelJasper:2010a}. In particular, the cubic term $\lambda \cos(\phi_1+\phi_2+\phi_3)$ constrains the total phase sum, whereas the bilinear term $u \cos(\phi_i-\phi_j)$ governs the relative phase differences between components. The observed $2\pi/3$ phase shift can be naturally captured by the combined action of these two terms (A detailed derivation is provided in the Supplementary Materials D). 

\begin{figure*}[tbp]
  \centering
  \includegraphics[width=0.8\textwidth]{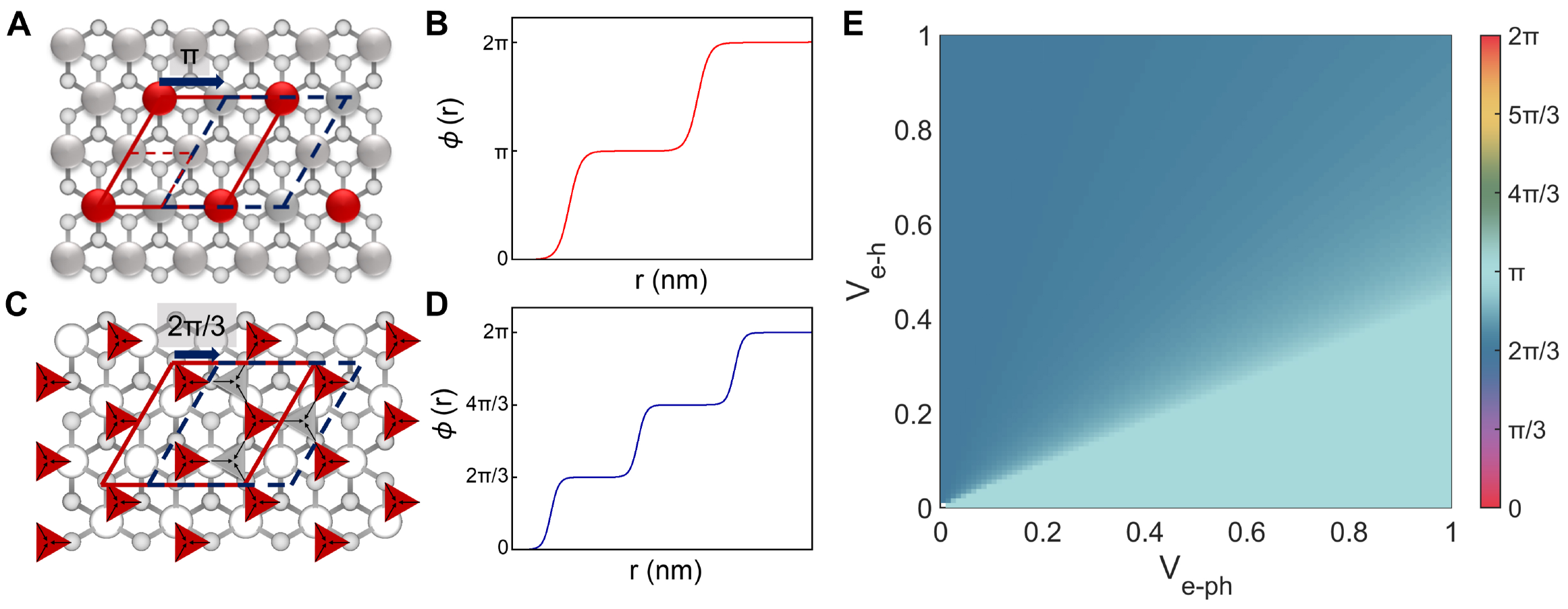}
  \caption{
  \textbf{Mechanism of the phase shift across the domain walls.}
  (A,B) Electron--phonon-driven scenario. The $2 \times 2$ supermodulation is represented by the red and gray Ti sites. A $\pi$ phase shift across the domain wall is indicated by the dashed parallelogram, with the corresponding phase profile $\phi(r)$ shown in (B).(C,D) Excitonic-driven scenario. The charge-transfer configuration is illustrated by the red triangles (upper Se-Ti-lower Se). The lowest-energy domain-wall configuration is obtained by transforming the red triangle into the gray one, yielding a $2\pi/3$ phase shift.(E) Phase diagram of the domain-wall phase shift as a function of $V_{e\text{-}h}$ and $V_{e\text{-}ph}$.}
  \label{fig:mechanism}
\end{figure*}

The phase configuration identified here is one of six symmetry-related permutations of the triple-$\mathbf{Q}$ state [i.e.$(\phi_1,\phi_2,\phi_3)$ = ($\pm 2\pi/3$, 0, 0),  (0, $\pm 2\pi/3$, 0) or (0, 0, $\pm 2\pi/3$)], characterized by relative phases of $0$ and $\pm 2\pi/3$. Although this configuration preserves the time-reversal symmetry of the real-valued charge modulation, it breaks the permutation symmetry among the three CDW components and therefore defines a distinct uniform reference state of the CDW order. Similar phase configurations have also been reported in 2\textit{H}-NbSe$_2$ \cite{LiuXiaolong:2021,ZhangHao:2025,CaoLu:2024}, suggesting that such inter-component phase locking may be a more general feature of transition metal dichalcogenides.

Moving to the domain-wall regions, two key questions arise: whether the inter-component couplings that stabilize the uniform phase-locked state remain dominant within each domain, and how the free-energy balance is modified at the domain boundary. The first issue is straightforward to answer. As shown in Fig.~\ref{fig:phase_cdw}, phase locking persists across the domain-wall region, but the global phase constraint is no longer strictly enforced locally. While one domain exhibits $\phi_1+\phi_2+\phi_3 = 4\pi/3$, the neighboring domain evolves toward a phase sum close to $0$. This indicates that the balance between the bilinear and cubic coupling terms becomes modified near the domain boundary, likely due to local strain \cite{LeiNa:2013,NatafGildasF:2020}, defects \cite{LiuXiaolong:2021,NatafGildasF:2020,CatalanGustau:2012}, or related perturbations. To address the second issue, we first note that the domain walls associated with the three CDW components are not spatially coincident, but instead appear at different locations (see Fig.S7 in Supplementary Materials H). Such weak spatial locking suggests that the lock-in energy acts primarily on each unidirectional CDW component rather than rigidly binding all three components together. The existence of spatially separated domain walls is therefore consistent with including a component-resolved lock-in term in the free-energy description.

We note that the energetics of domain-wall formation depend qualitatively on the microscopic origin of the CDW. In a lattice-driven scenario, the CDW is directly locked to the underlying crystal through the McMillan commensurability term $-V_{e\text{-}ph}\cos(2\phi_j)$. For the commensurate $2\times2$ CDW in 1\textit{T}-TiSe$_2$, this term favors phase minima separated by $\pi$, and therefore predicts a $\pi$ phase shift across the domain wall (Fig.~\ref{fig:mechanism}A and B). By contrast, in an excitonic scenario, the phase locking follows a different symmetry. Within the charge-transfer picture of exciton formation, an exciton-active center (blue triangle) can be identified, as illustrated in Fig.~\ref{fig:mechanism}C and proposed previously in Ref.~\cite{DiSalvoFrankJ:1976,vanWezelJasper:2010a}. In this picture, charge is transferred from a Ti atom to two neighboring Se atoms. A low-energy phase slip can then be realized by shifting the exciton-active center to a neighboring Se-Ti-Se triangle (gray triangle), as shown in Fig.~\ref{fig:mechanism}C and D. This process can naturally give rise to a $2\pi/3$ phase shift, motivating the inclusion of a higher-order lock-in term of the form $-V_{e-h}\cos(3\phi_j)$.

To capture the combined effects of lattice and electronic instabilities, as suggested by recent ultrafast experiments \cite{KurtzFelix:2024,LianChao:2020,HuberMaximilian:2024}, we write the phase-dependent lock-in free energy as
\begin{equation}
\begin{aligned}
\label{eq:GL_phase}
\mathcal{F}_{\mathrm{lock}}
=
\int dx \sum_{j=1}^{3}
&\Bigg[
\frac{\rho}{2}\left(\partial_x \phi_j\right)^2
- V_{e\text{-}ph}\cos(2\phi_j)\\
&- V_{e\text{-}h}\cos(3\phi_j)
\Bigg].
\end{aligned}
\end{equation}

\noindent Here, $\rho \equiv KA^2$, $V_{e\text{-}ph} \equiv V_0A^2$, and $V_{e\text{-}h} \equiv V_1A^3$, where $K$, $V_0$, and $V_1$ are phenomenological constants, and $A$ is the CDW amplitude. In this form, $V_{e\text{-}ph}$ and $V_{e\text{-}h}$ represent the effective strengths of the electron-phonon and electron-hole phase-locking contributions, respectively. Minimizing $\mathcal{F}_{\mathrm{lock}}$ yields the phase-shift solutions across the domain wall (see Supplementary Materials E for details). By tuning $V_{e\text{-}ph}$ and $V_{e\text{-}h}$, we map the resulting phase shift across the domain wall, as shown in Fig.~\ref{fig:mechanism}E. A clear boundary emerges at $V_{e\text{-}h}/V_{e\text{-}ph} \approx 0.45$. Below this boundary, the system favors a $\pi$ phase shift, whereas above it, the energetically preferred solution becomes $2\pi/3$. This result indicates that a purely lattice-driven description is insufficient to account for the experimentally observed $2\pi/3$ phase shift, implying that additional electronic interactions must be involved.

More broadly, our results establish the domain-wall phase structure as a real-space probe that constrains the microscopic interactions governing collective electronic order. Rather than identifying a unique driving mechanism, this work demonstrates that phase-resolved STM can access the internal phase organization of multi-component order parameters, an aspect that is often inaccessible to conventional momentum-space techniques. By connecting nanoscale phase textures to the underlying free-energy landscape, our approach provides a powerful framework for constraining competing and cooperative interactions in complex quantum materials. This capability opens new opportunities for investigating emergent symmetry breaking, topological defects, and interaction-driven phase competition in a wide range of correlated systems.

\section{Methods} 
\noindent\textbf{Samples:}  \\
1\textit{T}-TiSe$_2$ single crystals were grown by chemical vapor transport using iodine as the transport agent. High-purity titanium bar (99.99\%) and selenium lump (99.999\%) were weighed in a molar ratio of Ti:Se = 1:2.05, combined with iodine (5 mg$\cdot$cm$^{-3}$), and sealed under high vacuum (10$^{-4}$ Pa) in a pre-cleaned quartz ampoule. The loaded ampoule was positioned in a horizontal two-zone tube furnace, establishing a 100\,°C temperature gradient between the source zone (580\,°C) and the growth zone (680\,°C). Crystal growth proceeded over 21 days under these steady-state thermal conditions. Finally, the system was cooled to room temperature at a controlled rate of 10\,°C$\cdot$h$^{-1}$, yielding well-developed, layered single crystals exhibiting metallic luster.\\

\noindent\textbf{STM:}  \\
Scanning tunneling microscopy (STM) measurements were performed on single-crystalline 1\textit{T}-TiSe$_2$ samples using USM-1500 and USM-1300 STM systems operated under ultra-high vacuum (UHV) conditions with a base pressure better than $4.5 \times 10^{-10}$~Torr and base temperature better than 4.5~K. The samples were cleaved in situ at room temperature under UHV conditions and immediately transferred to the STM measurement stage to ensure a clean and well-defined surface. Tungsten tips were used throughout the experiments and were further conditioned in situ prior to measurements to achieve stable tunneling conditions. STM topographic images were acquired in constant-current mode. The tunneling current setpoint was adjusted accordingly to ensure optimal imaging stability. \\

\noindent\textbf{Data Availability}\\
All relevant data that support the findings of this study are presented in the manuscript and supplementary information file. All data are available upon reasonable request from the corresponding authors.\\

\noindent\textbf{Acknowledgments}\\
This work is supported by the National Key Basic Research Program of China
(Grants No. 2024YFF0727103). This work was supported by the Synergetic Extreme Condition User Facility (SECUF, https://cstr.cn/31123.02.SECUF). \\

\noindent\textbf{Author contributions}\\
G.H., J.X.Z., and J. Shen conceived the project. G.H., H.T.Z. and Z.H.S. performed the STM measurements. J.X.Z. and Z.X.W. contributed to the experimental assistance. G.H., H.T.Z., J.S\rev{hu}, J.L., Z.H.S., Z. M. L., Z.Y.D. and J.X.Z. analyzed the STM/S data. Z.C.X. and Y.G.S. synthesized and characterized the samples. G.H. and H.T.Z. wrote the manuscript with comments from all the authors.\\

\noindent\textbf{Competing interests}\\
The authors declare no competing interests.

\bibliography{refs}

%%%%%%%%%%%%%%%%%%%%%%%%%%%%%%%%%%%%%
% Supplementary Materials
%%%%%%%%%%%%%%%%%%%%%%%%%%%%%%%%%%%%%

\clearpage
\onecolumngrid

\section*{Supplementary Materials}

% Figure numbering
\setcounter{figure}{0}
\renewcommand{\thefigure}{S\arabic{figure}}

% Equation numbering
\setcounter{equation}{0}
\renewcommand{\theequation}{S\arabic{equation}}

\noindent \textbf{A. Two-dimensional lock-in analysis of CDW modulation}
~\\

\noindent \textbf{1) Lawler–Fujita correction}

All STM topographic maps $T_0(\mathbf{r})$ were first subjected to standard pre-processing procedures to remove extrinsic background contributions. A linear background was subtracted to eliminate large-scale height variations to suppress scan-line artifacts, yielding a flattened map $T(\mathbf{r})$ . To further correct slow spatial distortions arising from piezo creep and thermal drift, a lattice registration procedure based on the Lawler–Fujita (LF) algorithm was employed \cite{LawlerMJ:2010,FujitaKazuhiro:2014,LiuChengYen:2024}. We utilized the un-reconstructed atomic Bragg peaks $\mathbf{G}_i$ as an absolute structural reference to extract the spatial phase shift field $\theta_i(\mathbf{r})$ of the atomic lattice. The deviation from a perfect lattice is quantified by a continuous displacement field $\mathbf{u}(\mathbf{r})$:
\begin{equation}
\mathbf{u}(\mathbf{r}) = \begin{pmatrix} G_{1x} & G_{1y} \\ G_{2x} & G_{2y} \end{pmatrix}^{-1} \begin{pmatrix} \theta_1 - \theta_1(\mathbf{r}) \\ \theta_2 - \theta_2(\mathbf{r}) \end{pmatrix}.
\end{equation}
where $\theta_1$ and $\theta_2$ are global reference phases defining a uniform lattice gauge. The drift-corrected, geometrically perfect map is then obtained as
\begin{equation}
A(\mathbf{r}) = T(\mathbf{r} - \mathbf{u}(\mathbf{r})).
\end{equation}
This step ensures that the atomic lattice is spatially uniform and that the extracted CDW phase reflects intrinsic electronic modulations rather than instrumental distortions. The choice of global reference phases does not affect the relative CDW phase differences discussed in this work.\\

\noindent \textbf{2) Two-dimensional lock-in analysis}

The CDW wave vectors $\mathbf{Q}_i$ were identified from the Fourier transform $\tilde{A}(\mathbf{q})$ of the drift-corrected real-space data, where sharp satellite peaks corresponding to the triple-$\mathbf{Q}$ order are clearly resolved. To isolate each CDW component, a Gaussian window centered at $\mathbf{Q}_i$ was applied in momentum space:
\begin{equation}
\tilde{A}_{\mathbf{Q}_i}(\mathbf{q}) = \tilde{A}(\mathbf{q}) \cdot W_i(\mathbf{q}),
\end{equation}
where
\begin{equation}
W_i(\mathbf{q}) = \exp\left[-\frac{|\mathbf{q}-\mathbf{Q}_i|^2}{2\sigma_q^2}\right].
\end{equation}
This filtering suppresses contributions from neighboring wave vectors and high-frequency noise, while preserving the local CDW modulation associated with $\mathbf{Q}_i$. The width $\sigma_q$ determines the trade-off between spatial resolution and momentum selectivity, and was chosen such that the CDW peaks are well isolated while retaining the spatial variations across domain walls.

The spatially resolved CDW component at wave vector $\mathbf{Q}_i$ is obtained via a two-dimensional lock-in procedure, which can be expressed as a local convolution:
\begin{equation}
A_{\mathbf{Q}_i}(\mathbf{r}) = \int A(\mathbf{R}) \, e^{i \mathbf{Q}_i \cdot \mathbf{R}} \, G(\mathbf{R}-\mathbf{r}) \, d\mathbf{R},
\end{equation}
where $G(\mathbf{R}-\mathbf{r}) = \exp[-|\mathbf{R}-\mathbf{r}|^2/(2\sigma_r^2)]$ is a Gaussian window centered at $\mathbf{r}$. This operation assumes that the amplitude and phase vary slowly on the scale of the CDW wavelength, allowing a separation between the fast oscillatory component and the slowly varying envelope.

For computational efficiency, this procedure is implemented in momentum space as:
\begin{equation}
A_{\mathbf{Q}_i}(\mathbf{r}) = \mathcal{F}^{-1} \left\{ \mathcal{F}\left[A(\mathbf{r}) e^{i\mathbf{Q}_i \cdot \mathbf{r}}\right] \cdot \hat{G}(\mathbf{q}) \right\}.
\end{equation}

\noindent The resulting complex field can be written as
\begin{equation}
A_{\mathbf{Q}_i}(\mathbf{r}) = A_i(\mathbf{r}) e^{i\phi_i(\mathbf{r})},
\end{equation}
from which the local amplitude and phase are obtained as
\begin{equation}
A_i(\mathbf{r}) = \sqrt{\mathrm{Re}[A_{\mathbf{Q}_i}]^2 + \mathrm{Im}[A_{\mathbf{Q}_i}]^2},
\end{equation}
\begin{equation}
\phi_i(\mathbf{r}) = \mathrm{atan2}(\mathrm{Im}[A_{\mathbf{Q}_i}], \mathrm{Re}[A_{\mathbf{Q}_i}]).
\end{equation}

The amplitude $A_i(\mathbf{r})$ reflects the local strength of the CDW modulation, while the phase $\phi_i(\mathbf{r})$ encodes its spatial configuration. Uniform regions correspond to phase-coherent domains, whereas spatial variations in $\phi_i(\mathbf{r})$ directly reveal phase textures such as domain walls.

To ensure robustness, we verified that the extracted phase patterns are insensitive to moderate variations of the Gaussian window size. Phase unwrapping was performed where necessary to ensure continuity over extended regions, while all reported phase differences are defined modulo $2\pi$. We further confirmed that the observed phase shifts across domain walls are not affected by residual drift correction or filtering procedures, indicating their intrinsic origin. Consistency among the three symmetry-related CDW components was independently verified, ensuring that the extracted phase relations reflect intrinsic multi-$\mathbf{Q}$ coupling rather than analysis artifacts.\\

\noindent \textbf{B. Reproducibility of the Phase configuration in  uniform regions}
~\\

\begin{figure*}[bp]
    \centering
    \includegraphics[width=0.9\textwidth]{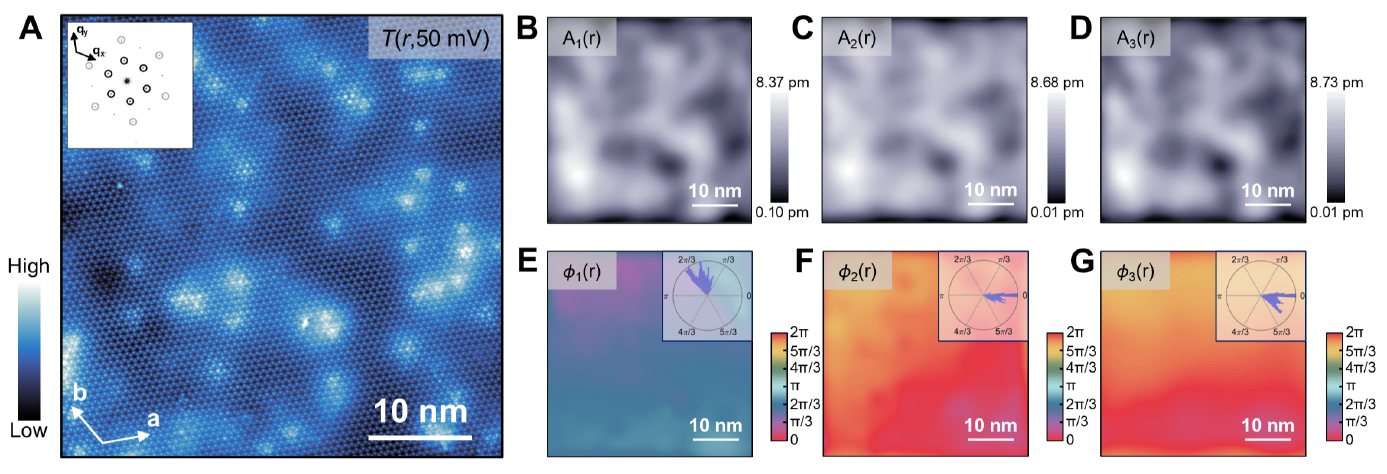}
    \caption{
    \textbf{Phase-resolved STM characterization of a uniform triple-$\mathbf{Q}$ CDW region.}
(A) STM topographic image acquired over a $45\,\mathrm{nm} \times 45\,\mathrm{nm}$ field of view ($V = 50$~mV, $I = 1$~nA), showing a uniform CDW modulation without visible defects or domain boundaries. Upper-left inset: Two-dimensional FFT of (A).
(B--D) Amplitude maps $A_i(\mathbf{r})$ obtained from 2D lock-in analysis, showing spatially uniform CDW intensity.
(E--G) Corresponding phase maps $\phi_i(\mathbf{r})$. Insets: Histograms of the phase values, with peaks near $2\pi/3$, $0$, and $0$ (mod $2\pi$).
    }
    \label{fig:stm_uniform}
\end{figure*}

Figure~\ref{fig:stm_uniform} presents phase-resolved STM measurements obtained in a spatially uniform region distinct from that shown in the main text. The STM topographic image (Fig.~\ref{fig:stm_uniform}A) displays a well-ordered CDW modulation without visible defects or domain boundaries. The corresponding FFT (upper-left inset of Fig.~\ref{fig:stm_uniform}A) exhibits sharp Bragg peaks and well-defined CDW satellite peaks, confirming the presence of a robust commensurate triple-$\mathbf{Q}$ order. The amplitude maps associated with the three symmetry-equivalent CDW wave vectors (Figs.~\ref{fig:stm_uniform}(B-D)) are spatially homogeneous across the entire field of view, with no detectable suppression or discontinuity. This confirms the absence of domain walls and indicates that the system resides in a single, uniform CDW domain.The corresponding phase maps (Figs.~\ref{fig:stm_uniform}(E-G)) exhibit nearly constant phase values, with only weak long-wavelength modulations. Phase histograms [insets of Figs.~\ref{fig:stm_uniform}(E--G)] show that one component is centered near $2\pi/3$, while the other two are clustered around $0$ ($2\pi$). Taking into account the $2\pi$ periodicity, the phase configuration can be expressed as $(\phi_1,\phi_2,\phi_3) \approx (2\pi/3,0,0)$.

This observation establishes an alternative uniform reference state of the triple-$Q$ CDW, which is symmetry-related to the phase configuration identified in the main text. The reproducibility of such fractionally phase-locked states across independent regions demonstrates that the observed phase relation is an intrinsic property of the CDW order.\\

\noindent \textbf{C. Reproducibility of phase evolution in domain regions}
~\\

\begin{figure*}[bp]
    \centering
    \includegraphics[width=0.8\textwidth]{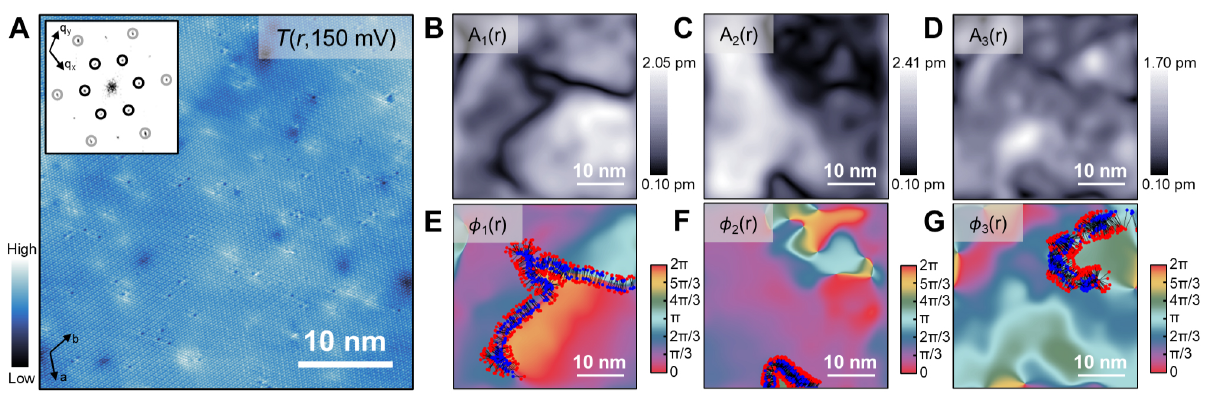}
    \caption{
    \textbf{Phase-resolved STM characterization of CDW domain walls in an independent region.}
(A) STM topographic image acquired over a $40\,\mathrm{nm} \times 40\,\mathrm{nm}$ field of view ($V = 150$~mV, $I = 300$~pA). Inset: FFT showing lattice Bragg peaks and CDW satellite peaks.
(B-D) Amplitude maps $A_i(\mathbf{r})$ revealing spatially extended domain-wall regions with suppressed CDW amplitude.
(E-G) Corresponding phase maps $\phi_i(\mathbf{r})$. Phase differences across the domain walls are evaluated by selecting pairs of sampling points on opposite sides of the domain wall and computing the phase difference $\Delta\phi$ between them.
    }
    \label{fig:stm_nonuniform_domainwalls}
\end{figure*}

\begin{figure}[tbp]
    \centering
    \includegraphics[width=0.8\columnwidth]{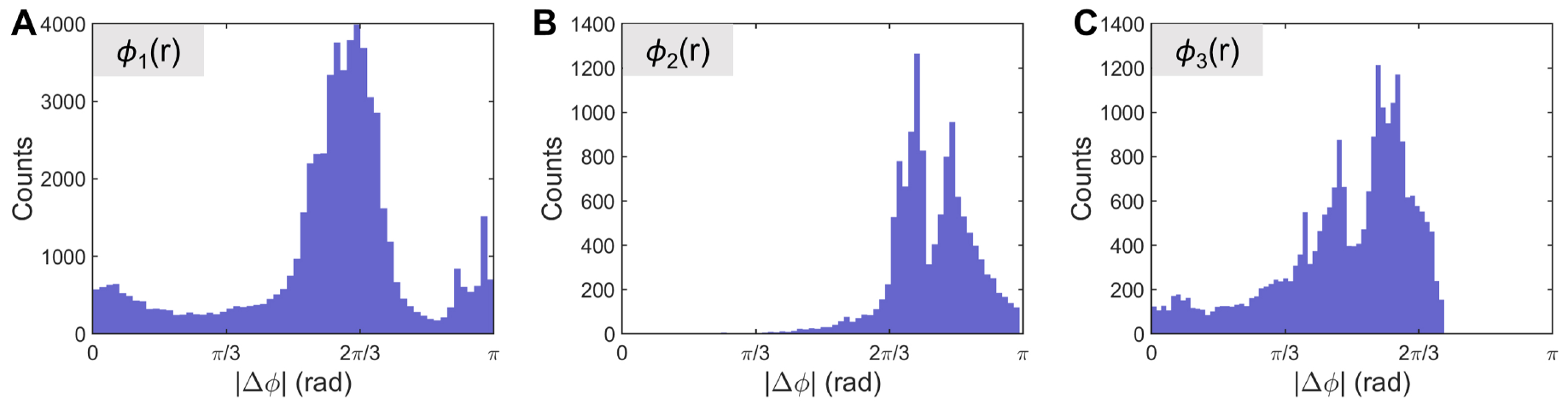}
    \caption{
    \textbf{Statistical analysis of phase differences across domain walls in an independent region.}
(A-C) Histograms of the phase difference $\Delta\phi$ for the three CDW components, extracted across the domain walls shown in Fig.~\ref{fig:stm_nonuniform_domainwalls}(E-G). The phase differences are obtained using the same sampling procedure as described in the main text.
    }
    \label{fig:stm_nonuniform_hist}
\end{figure}

To further validate the domain-wall phenomenology discussed in the main text, we analyze an additional non-uniform region, shown in Fig.~\ref{fig:stm_nonuniform_domainwalls}, using the same domain-wall-based phase-difference extraction procedure. Fig.~\ref{fig:stm_nonuniform_domainwalls}(A) displays an STM topographic image, with the corresponding FFT confirming the presence of long-range triple-$Q$ CDW order. The amplitude maps (Fig.~\ref{fig:stm_nonuniform_domainwalls}(B-D)) reveal pronounced domain-wall features. As in the main text, the amplitude suppression associated with each CDW component occurs at distinct spatial locations, indicating that the domain walls of the three components are not spatially coincident. The phase maps (Fig.~\ref{fig:stm_nonuniform_domainwalls}(E-G)) exhibit smooth spatial evolution across the domain walls. To quantify the phase variation, we evaluate the phase difference $|\Delta\phi|$ by selecting pairs of sampling points located on opposite sides of the domain wall, following the same procedure described in the main text. This approach provides a local measure of the phase discontinuity across the wall. The resulting statistical distributions of $|\Delta\phi|$ for the three CDW components are summarized in Fig.~\ref{fig:stm_nonuniform_hist}. In all cases, the phase differences exhibit well-defined peaks centered near $2\pi/3$, with relatively narrow distributions. This behavior is robust against variations in local domain-wall geometry and spatial location.

These observations reproduce the key features identified in the main text, namely the spatial separation of domain walls among different CDW components and a characteristic fractional phase shift. The consistency across independent regions further supports that the observed domain-wall phase structure is an intrinsic property of the triple-$Q$ CDW state.\\

\noindent \textbf{D. Ginzburg--Landau analysis of the uniform phase configuration}
~\\

To provide a quantitative foundation for the uniform phase configuration discussed in the main text, we present here a detailed derivation of the phase-dependent Ginzburg--Landau (GL) free energy and its minimization.

The triple-$Q$ CDW is described by three complex order parameters\cite{vanWezelJasper:2010a,MonneyClaude:2011}
\begin{equation}
\Psi_j(\mathbf{r}) = A_j(\mathbf{r}) e^{i\phi_j(\mathbf{r})}, \quad j=1,2,3,
\end{equation}
associated with wave vectors $\mathbf{Q}_j$ satisfying $\mathbf{Q}_1+\mathbf{Q}_2+\mathbf{Q}_3=0$.

The GL free energy must respect both lattice symmetry and translational invariance.\cite{McMillanWilliamL:1976} In particular, phase-dependent coupling terms are constrained by momentum conservation modulo reciprocal lattice vectors. To lowest order, two independent inter-component phase couplings are allowed\cite{NakanishiKazuo:1977,vanWezelJasper:2010a}.

The first is a bilinear coupling of the form
\begin{equation}
u \sum_{i<j} |\Psi_i \Psi_j|,
\end{equation}
which originates from pairwise interactions between CDW components. Upon expressing $\Psi_j = A_j e^{i\phi_j}$, this term gives rise to a phase-dependent contribution proportional to $\cos(\phi_i - \phi_j)$, and therefore governs the relative phase differences between components.

The second is a cubic coupling term
\begin{equation}
\lambda |\Psi_1 \Psi_2 \Psi_3|,
\end{equation}
which is symmetry-allowed because $\mathbf{Q}_1+\mathbf{Q}_2+\mathbf{Q}_3=0$. This term couples all three components simultaneously and depends on the total phase $\phi_1+\phi_2+\phi_3$.

These two terms constitute the lowest-order symmetry-allowed contributions that couple the phases of different CDW components.

In the uniform region identified in Fig.~2 of the main text, both amplitudes and phases are approximately constant over the STM field of view, and spatial gradients can be neglected. The amplitudes may therefore be taken as equal, $A_j = A$. Substituting $\Psi_j = A e^{i\phi_j}$ into the GL free energy and retaining only the phase-dependent terms yields\cite{McMillanWilliamL:1976}
\begin{equation}
\mathcal{F}_{\mathrm{phase}} =
2\lambda A^3 \cos(\phi_1 + \phi_2 + \phi_3)
+ u A^2 \sum_{i<j} \cos(\phi_i - \phi_j).
\end{equation}

Here, we consider a representative parametrization $\phi_1=0$, $\phi_2=\varphi$, and $\phi_3=0$. The free energy reduces to
\begin{equation}
\mathcal{F}_{\mathrm{phase}}(\varphi)
=
2\lambda A^3 \cos\varphi
+
u A^2 (1 + 2\cos\varphi).
\end{equation}

Minimization with respect to $\varphi$ yields
\begin{equation}
\sin\varphi \left( \lambda A + u \cos\varphi \right) = 0.
\end{equation}

In addition to the trivial solutions $\varphi=0$ and $\pi$, a nontrivial solution exists when
\begin{equation}
\cos\varphi = -\frac{\lambda A}{u}.
\end{equation}

For the experimentally observed value $\varphi=4\pi/3$, corresponding to $\cos\varphi=-1/2$, this condition requires $\lambda A = u/2$. The second derivative confirms that this solution corresponds to a local minimum for $u>0$.

The above analysis demonstrates that the experimentally observed fractional phase configuration arises from the cooperative interplay between the two symmetry-allowed phase coupling terms. The bilinear term $u \cos(\phi_i-\phi_j)$ determines the relative phase differences between CDW components, while the cubic term $\lambda \cos(\phi_1+\phi_2+\phi_3)$ imposes a global constraint on the total phase. Neither term alone can stabilize a $2\pi/3$ phase shift: the bilinear term alone favors alignment or $\pi$-type offsets, while the cubic term alone constrains only the phase sum without fixing relative phase differences. Only when both terms are present does the system admit a stable solution with fractional phase locking, consistent with the experimental observation. The configuration $(0,4\pi/3,0)$ is one of six symmetry-related minima characterized by relative phase differences of $0$ and $\pm 2\pi/3$, forming a discrete manifold of degenerate uniform states\cite{IshiokaJunya:2010}.\\

\noindent \textbf{E. GL analysis of phase difference across the domain wall with competing mechanisms}
~\\

Assuming that the amplitude of the CDW order parameter is rigid, we focus on the phase degree of freedom and adopt a phase-only Ginzburg-Landau free energy:

\begin{equation}
\mathcal{F}_{\text{lock}} = \int dx \left\{ 
\frac{\rho}{2} (\partial_x \theta)^2 
- V_{e\text{-}ph} \cos(2\theta) 
- V_{e\text{-}h} \cos(3\theta) 
\right\}.
\end{equation}

Here, $\theta(x)$ is the phase field and $\rho$ is the phase stiffness. 
The cosine terms represent commensurate lock-in potentials 
originating from different microscopic mechanisms. 
The $m=2$ term ($\cos 2\theta$) typically arises from  electron-phonon coupling\cite{WeberFrank:2011,HoltMartin:2001,OttoMartinR:2021,KaramTonyE:2018}, while the $m=3$ term ($\cos 3\theta$) can originate from 
purely electronic effects such as electron-hole interactions\cite{KiddThomasE:2002,CercellierHerve:2007,SugawaraKatsuaki:2016,KogarAnshul:2017}. The competition between these terms leads to multiple local minima in the phase potential, which can be given by the Euler--Lagrange equation,

\vspace{0.5em}
\noindent
\begin{equation}
\rho \, \theta''(x) 
= 2V_{e\text{-}ph} \sin(2\theta) 
+ 3V_{e\text{-}h} \sin(3\theta).
\label{EL}
\end{equation}

Multiplying Eq.\ref{EL} by $\theta'(x)$ and integrating $dx$ yields
\begin{equation}
\frac{\rho}{2} (\theta')^2 = U(\theta) + E_0.
\end{equation}

\noindent where $U(\theta) = - V_{e\text{-}ph} \cos(2\theta) - V_{e\text{-}h} \cos(3\theta)$ and $E_0$ is a constant energy. For a domain wall connecting two minima, the boundary condition
\begin{equation}
\theta'(x \to \pm \infty) = 0
\end{equation}
Sets $E_0 = -U_{\mathrm{global~min}}$, leading to
\begin{equation}
\frac{\rho}{2} (\theta')^2 = U(\theta) - U_{\mathrm{global~min}}.
\end{equation}

%\textbf{(ii) Implicit spatial solution}

\begin{figure}[h]
\centering
\includegraphics[width=1\linewidth]{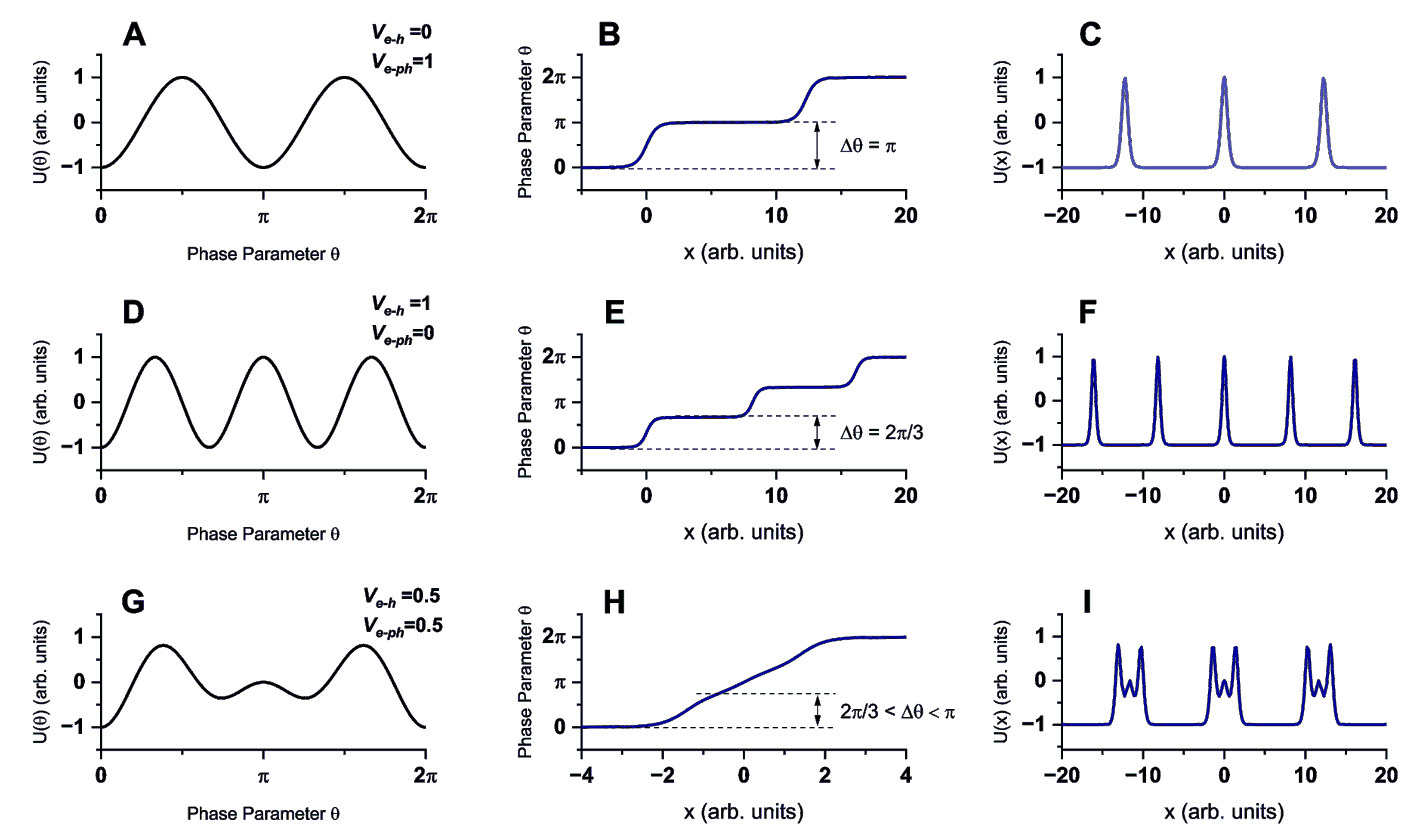}
\caption{
    \textbf{Phase evolution for three representative parameter sets.}
(A,D,G) the potential $U(\theta)$, (B,E,H) the corresponding phase profile $\theta(x)$, and (C,F,I) the resulting energy density $U(x)=U[\theta(x)]$.
    }
\label{fig:placeholder}
\end{figure}

This equation defines a relation between the phase $\theta$ and the spatial coordinate $x$:
\begin{equation}
\label{fc: x-theta}
x - x_0 = \int \sqrt{\frac{\rho}{2}}
\frac{d\theta}{\sqrt{U(\theta) - U_{\mathrm{global~min}}}}.
\end{equation}

\noindent It describes a smooth trajectory connecting the global minima at $x \to \pm \infty$, fully determining the domain-wall structure in an infinite system. The resulting phase profiles and corresponding energy densities for representative parameter sets are shown in Fig.~\ref{fig:placeholder}.

In a finite system, however, this constraint is relaxed, and the phase field may occupy nearby local minima over a finite spatial region. The energetic cost of such a configuration is proportional to the excess potential relative to the global minimum multiplied by the region size. If this energy remains small compared to the gradient term, the solution can still be treated as a perturbation of the infinite-system trajectory. In this perspective, the spatial profile can be approximated by
\begin{equation}
\label{fc: x-theta2}
x - x_0 = \int \sqrt{\frac{\rho}{2}}
\frac{d\theta}{\sqrt{U(\theta) - U_{\mathrm{local~min}}}}.
\end{equation}
Each local minimum with positive curvature $U''(\theta) > 0$, defines a locally stable phase region. Although the integral diverges at infinity, in practice $\theta(x)$ approaches the local minimum exponentially:
\begin{equation}
\theta(x) - \theta_i \sim e^{-|x-x_0|/\xi_i},
\end{equation}
so that over a finite spatial extent, the solution already appears effectively pinned, forming a plateau whose width increases as the energy approaches the local minimum.

\begin{figure}[h]
\centering
\includegraphics[width=1\linewidth]{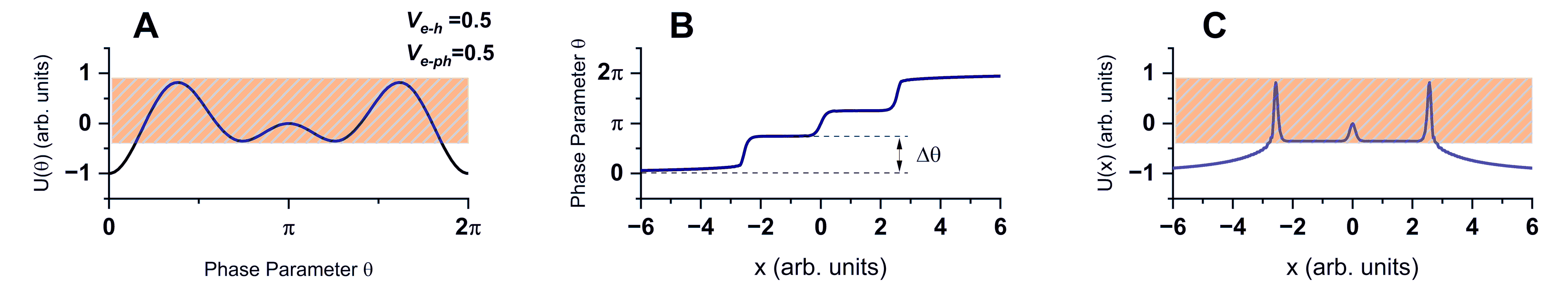}
\caption{
\textbf{Finite-size phase evolution for $(V_{e\text{-}ph}, V_{e\text{-}h}) = (0.5,0.5)$.}
(A) the potential $U(\theta)$, (B) the phase profile $\theta(x)$ obtained from Eq.~(\ref{fc: x-theta2}), and (C) the corresponding spatial energy density $U(x)$. The shaded regions highlight the domain where the solution remains real and physically relevant.
    }
\label{fig:placeholder2}
\end{figure}

This behavior is illustrated in Fig.~\ref{fig:placeholder2}, where the shaded regions in (A) and (C) indicate the domain with $U(\theta) \ge U_{\mathrm{local~min}}$, within which the solution remains physically accessible.

Thus, although strictly infinite in theory, these solutions provide a physically meaningful approximation for finite systems, justifying the presence of locally stabilized phase regions associated with metastable minima. The physically relevant phases correspond to the \textbf{local minima}:
\begin{equation}
\theta_i = \arg\min_{\theta \in \mathcal{N}(\theta^*)} U(\theta) .
\end{equation}

In the limiting cases where only a single term is present, the minima are uniformly spaced: $\theta = 0, \pi, 2\pi$ for the $\cos(2\theta)$ term and $\theta = 0, 2\pi/3, 4\pi/3, 2\pi$ for the $\cos(3\theta)$ term. In both cases, $\theta = 0$ remains a common reference minimum. When both terms are present, the positions of the minima evolve continuously. Taking $\theta_0 = 0$ as a reference, we define the phase difference to the nearest adjacent minimum as
\begin{equation}
\Delta\theta = \theta_1 - \theta_0 = \theta_1.
\end{equation}

Within the finite-size framework described above, where all local minima are admissible, $\Delta\theta$ provides a compact measure of the effective phase spacing selected by the competing terms. It interpolates between the characteristic values of the two limiting cases and therefore serves as a convenient parameter for mapping out the phase diagram in the $(V_{e\text{-}ph}, V_{e\text{-}h})$ plane.\\

\begin{figure}[t]
\centering
\includegraphics[width=0.8\linewidth]{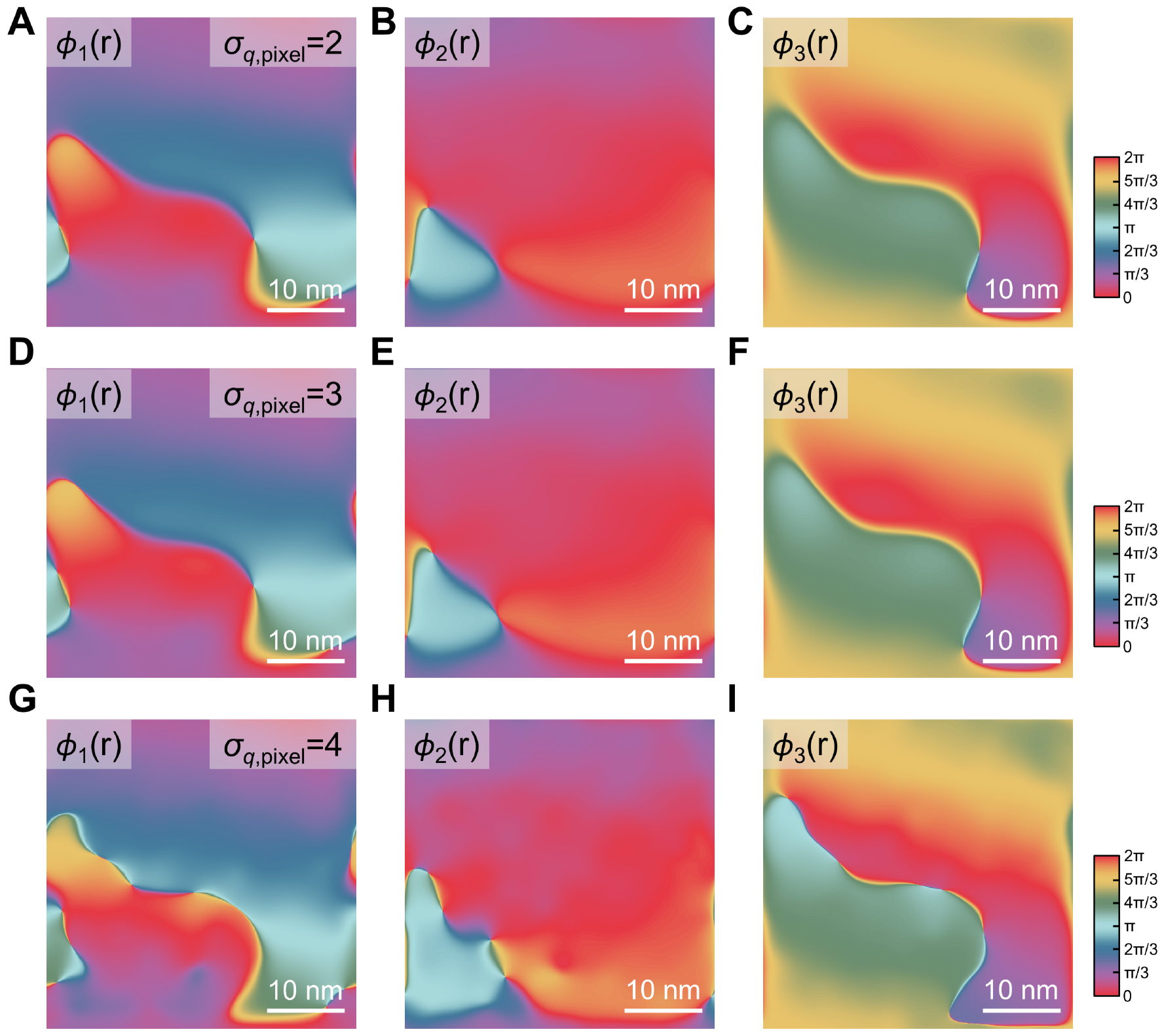}
\caption{
\textbf{Robustness of phase extraction against Gaussian filtering strength.}
Phase maps of the triple-$Q$ CDW obtained using different Gaussian filter parameters in the lock-in demodulation procedure. 
(A–C) $\sigma_{q,\mathrm{pixel}}=2$, (D–F) $\sigma_{q,\mathrm{pixel}}=3$ , and (G–I) $\sigma_{q,\mathrm{pixel}}=4$.
}
\label{fig:gaussian_robustness}
\end{figure}

\noindent \textbf{F. Real-Space Gaussian Broadening}
~\\

In the lock-in demodulation procedure\cite{FujitaKazuhiro:2014}, the Gaussian windows applied in momentum space and real space satisfy the relation
\begin{equation}
    \sigma_r \sigma_q = \frac{1}{2\pi},
\end{equation}
where $\sigma_r$ and $\sigma_q$ are the Gaussian widths in real space and momentum space, respectively.

However, in pixel units, the relation changes to
\begin{equation}
    \sigma_{r,\mathrm{pixel}} \sigma_{q,\mathrm{pixel}} = \frac{N}{2\pi},
\end{equation}
due to the scale invariance of Gaussian functions under the discrete Fourier transform.

The connection between the two representations is given by
\begin{equation}
    \sigma_{r,\mathrm{pixel}} = \sigma_r \frac{N}{L},
\end{equation}
and
\begin{equation}
    \sigma_{q,\mathrm{pixel}} = \sigma_q L.
\end{equation}

In the data analysis, we are interested in the relation between $\sigma_{q,\mathrm{pixel}}$ and $\sigma_r \equiv \sigma_r^{\mathrm{rel}} a_0$, where $\sigma_r^{\mathrm{rel}}$ is the Gaussian width in real space normalized by the lattice constant $a_0$. The relation is given by
\begin{equation}
    \sigma_r^{\mathrm{rel}} =
    \frac{L}{2\pi a_0 \sigma_{q,\mathrm{pixel}}}.
\end{equation}

Thus, the Gaussian width chosen in momentum space in pixel units can be directly related to the corresponding relative width in real space.\cite{LiuChengYen:2024}

The STM data analyzed in the main text, focusing on the domain-wall regions [see Fig.~2], correspond to a field of view of $L = 40$~nm sampled on an $N \times N = 1024 \times 1024$ grid, yielding a real-space pixel size of $\Delta r = L/N \approx 0.03906$~nm. For 1\textit{T}-TiSe$_2$, the in-plane lattice constant is $a_0 \approx 0.354$~nm. In this work, we set $\sigma_{q,\mathrm{pixel}} = 3.6$, which corresponds to a real-space width of approximately $5a_0$.

As shown in Fig.~2 of the main text, the domain-wall regions exhibit sharp spatial variations in the phase fields, which are smoothed by the Gaussian filtering, leading to an apparent broadening of the domain walls. However, the phase differences are evaluated deep inside uniform domains, at distances $d \gg \sigma_r$, where the phase fields are spatially homogeneous. Consequently, the extracted macroscopic phase values are largely insensitive to boundary-induced coarse-graining effects.

To further verify the robustness of the phase extraction, we systematically varied the filter parameter $\sigma_{q,\mathrm{pixel}}$ over a range of values ($\sigma_{q,\mathrm{pixel}} = 2, 3, 4$), corresponding to real-space Gaussian widths 9$a_0$, 6$a_0$ and 4.5$a_0$, respectively. As shown in Fig.~\ref{fig:gaussian_robustness}, increasing $\sigma_{q,\mathrm{pixel}}$ leads to progressively sharper domain-wall profiles, reflecting enhanced spatial coarse-graining. Nevertheless, the spatial arrangement of phase domains and the relative phase offsets remain unchanged. In particular, the phase difference between adjacent CDW domains remains locked at approximately $2\pi/3$ within numerical uncertainty for all values of $\sigma_{q,\mathrm{pixel}}$. These results demonstrate that the observed fractional phase shift is not sensitive to the specific choice of filtering parameter, but instead reflects a robust feature of the extracted phase field.\\

\noindent \textbf{G. Extraction of phase difference across domain walls}\\

To quantify the phase discontinuity across domain walls, we perform a domain-wall-centered real-space analysis based on the phase maps obtained from lock-in analysis. This procedure provides a direct measure of the local phase difference across individual domain walls and complements the analysis presented in the main text.

The domain wall locations are first identified from the amplitude maps of the individual CDW components. Regions with strongly suppressed amplitudes are defined as domain wall areas using a percentile-based threshold, typically corresponding to the lowest 10\% of the amplitude distribution. To minimize contributions from isolated defects and boundary artifacts, only the largest connected domain wall structure is retained for further analysis.

For each point along the domain wall, phase values are sampled from the two neighboring domains located on opposite sides of the wall. To improve statistical robustness, the phase is averaged over finite spatial regions surrounding the sampling locations. The separation between the two sampling regions is chosen to exceed the intrinsic domain wall width while ensuring that both regions remain within the adjacent domains. In practice, this distance is adjusted according to the local domain wall width to access representative phase values away from the domain wall core.

The phase difference across a domain wall is defined as the difference between the averaged phase values measured on the two sides of the wall, followed by a modulo-$2\pi$ operation to account for the periodicity of the phase. This procedure removes ambiguities associated with phase wrapping and provides a robust measure of the phase discontinuity across the domain wall. The resulting phase difference is independent of the choice of global phase reference and remains stable against moderate variations in the sampling distance.\\

\noindent \textbf{H. Identification of domain walls}
~\\

To determine the spatial locations of domain walls  associated with each CDW component, we analyze the amplitude maps obtained from the 2D lock-in procedure. As discussed in the main text, domain wall regions are characterized by a local suppression of the CDW amplitude.

For each component, domain walls are identified by applying a threshold to the corresponding amplitude map $A_i(\mathbf{r})$, selecting regions where the amplitude is significantly reduced relative to its typical value within the domain. To improve robustness against noise and small-scale fluctuations, the amplitude maps are first smoothed using a Gaussian filter prior to thresholding. The resulting domain wall regions are further refined by retaining only the dominant connected structures, thereby excluding isolated defects or spurious features.

\begin{figure*}[h!]
\centering
\includegraphics[width=0.4\columnwidth]{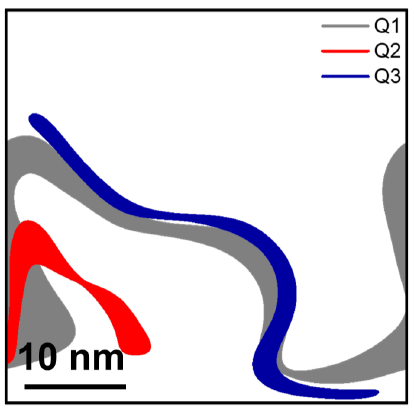}
\caption{
\textbf{Spatial identification of domain walls for the three CDW components.}
}
    \label{fig:dw_identification}
\end{figure*}

Figure~\ref{fig:dw_identification} shows the extracted domain wall locations for the three CDW components. The domain walls form extended line-like structures corresponding to regions of locally suppressed amplitude, but are clearly not spatially coincident across different components. Instead, each component exhibits a distinct domain wall network with different spatial positions, indicating component-resolved domain-wall formation.

This spatial separation of domain walls directly supports the observation discussed in the main text, demonstrating that the three CDW components are not rigidly locked to a common boundary. Rather, the domain walls are independently formed within each component, consistent with a weak spatial coupling between the different CDW modulations.

\end{document}